# Enzyme-Based Logic:
# OR Gate with Double-Sigmoid Filter Response

Oleksandr Zavalov,[a] Vera Bocharova,[b,c] Vladimir Privman,[a] Evgeny Katz[b]

[a] Department of Physics, Clarkson University, Potsdam, NY 13699, USA

[b] Department of Chemistry and Biomolecular Science, Potsdam, NY 13699, USA

[c] Oak Ridge National Laboratory, Oak Ridge, Tennessee 37831-6197, USA

## Abstract

The first realization of a biomolecular **OR** gate function with double-sigmoid response (sigmoid in both inputs) is reported. Two chemical inputs activate the enzymatic gate processes resulting in the output signal: chromogen oxidation, which occurs when either one of the inputs or both are present (corresponding to the **OR** binary function), and can be optically detected. High-quality gate functioning in handling of sources of noise is enabled by "filtering" involving pH control with an added buffer. The resulting gate response is sigmoid in both inputs when proper system parameters are chosen, and the gate properties are theoretically analyzed within a model devised to evaluate its noise-handling properties.

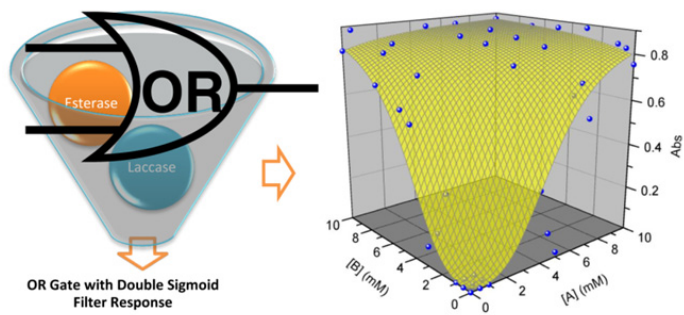

## INTRODUCTION

One of the promising approaches to novel biosensing and computing[1,2] for biotechnology has been the use of molecular[3,4] and biomolecular[5,6] systems for information processing. Biochemical[7-11] and chemical[12-16] processes have been utilized for various implementations as logic elements and networks for signal processing and novel computation designs. Biomolecular systems used in these applications have included those based on proteins/enzymes,[9,17] DNA,[7,18] RNA[19] and whole cells.[20,21] Applications could include the design of biosensors capable of processing and multiplexing several biochemical signals in the binary format, **0** and **1**. Biomolecular computing systems in biochemical and biotechnological environments[22] promise design of a new class of such sensors, with the information processing carried out by (bio)chemical processes rather than electronics.[23,24] Logic-mimicking processes based on biomolecular systems are being studied and developed[25-35] for potential applications in diagnostic and biomedical devices, including those responding to biomarkers characteristic of various pathophysiological conditions of diseases or injuries.

Since biological environments and enzymatic process kinetics are noisy, in practical situations it is challenging to discriminate between the signals or signal ranges designated as "binary" **0** or **1** for the YES/NO type diagnostics. The conceptual approach involving "digital processing" of biochemical signals aims at producing the output signal as well-defined YES/NO (**0** or **1**) values, allowing direct coupling of the signal processing with chemical actuators ultimately resulting in integrated biosensor-bioactuator "Sense/Act/Treat" systems.[23,24] Such digital biocomputing integrates biological and electronic/computing concepts, and requires a careful optimization as part of the gate and network designs.[25-29]

Optimization of biochemical reaction based gates for digital functioning can be accomplished by several approaches.[25-29,36-39] Recently, we have shown[28,36-39] that adding a (bio)chemical "filter" process allows changing a convex shape of the response function — which is typical for most of (bio)catalytic reactions — to a sigmoid shape. This approach solves the problem of discrimination of and noise control in the output signal, at least at the single-gate level, more efficiently than earlier-tried parameter tuning. Several (bio)chemical systems with added "filtering" have recently been designed[36-39] and optimized as standalone elements for inclusion in biochemical logic networks.

Most gate functions devised[5,6,9] for "biocomputing," mimic those of the two-input/one-output binary Boolean logic. This approach has been motivated by compatibility with electronic circuit/network designs and requires two-input gates. However, except for a single very recent result[25,29,40] for an **AND** gate, all the other reported "filtering" approaches[29,36-39] have thus far been realized for a single-input processing corresponding to transduction or "identity" gate. Furthermore, for shorter-term applications of biomolecular signal processing for diagnostics of certain types of injuries,[41,42] typically at least two markers provide the necessary level of the output signal — such as that of the **AND**[29,39] function — for a confident YES/NO determination of the specific medical condition, for instance, liver injury.[43,44] In the latter example, the added filtering has allowed not only to improve the process, but for the relevant measurement "gate times" actually enabled an accurate **AND** binary logic implementation.[39,45]

Therefore, it is important to incorporate and test various filtering approaches with a "toolbox" of binary gates, including **AND**, **OR**, etc. This paper presents the first realization of an **OR** gate with incorporated filtering. We utilize a biocatalytic reaction producing pH changes[46-50] with the pH-buffering mechanism[37] accomplished by adding the appropriately selected amount of a buffer. We carry out an experimental mapping of the response function of the realized **OR** gate. A numerical kinetic model is then utilized to quantify the behavior of the system and its noise-handling when it functions as an **OR** binary gate. We demonstrate that the added appropriately designed filtering step makes the gate's response sigmoid.

**EXPERIMENTAL**

*Chemicals and Materials*

Esterase from *Sus scrofa* (porcine liver) (E.C. 3.1.1.1), laccase from *Trametes versicolor* (E.C. 1.10.3.2), tris(hydroxymethyl)aminomethane (Tris buffer), potassium hexacyanoferrate (II) trihydrate and methyl butyrate 99% were purchased from Sigma-Aldrich. Ethyl butyrate 99% was purchased from Fluka. All commercial chemicals were used as supplied without further purification. Ultrapure water (18.2 MΩ·cm) from a NANOpure Diamond (Barnstead) source was used in all of the experiments.

*Composition and mapping of the **OR** gate with double sigmoid filter response*

The main process of the **OR** gate was based on the reaction biocatalyzed by esterase (0.5 U·mL$^{-1}$ equal to $4.85 \times 10^{-5}$ mM based on 67 kDa molecular weight[51]) converting Input 1, ethyl butyrate, or Input 2, methyl butyrate, or both of them to butyric acid and respective alcohols. The experiments were performed in a non-buffered solution of 50 mM sodium sulfate or in Tris buffer (4 mM or 8 mM) with the initial pH adjusted to 9.0. Because butyric acid produced by the hydrolytic activity of esterase partially dissociates in aqueous solution the initial pH of the solution decreases. It is known[52,53] that overall rate of reactions biocatalyzed by laccase (the second enzyme in the system) strongly depends on pH, being very slow at pH values above 8.1. Therefore, change in pH, caused by the hydrolytic reaction biocatalyzed by esterase, activates the reaction of laccase (0.5 U·mL$^{-1}$ equal to $3.78 \times 10^{-4}$ mM based on the average molecular weight of 66 kDa for fungal laccases[54-57]) which catalyzes the oxidation of 1 mM ferrocyanide in the presence of oxygen. The schematic illustration of the composition of the **OR** gate is represented in Scheme 1 (see page 17). In the gate-response-mapping experiments the input signals were applied at (separately) variable concentrations, ethyl butyrate: 0, 0.5, 1, 3, 5, 7 and 10 mM, and methyl butyrate: 0, 0.5, 1, 3, 5, 7 and 10 mM. There were thus 49 experimental data sets. Constant oxygen concentration during the experiment was maintained by bubbling air in the top part of the solution.

*Optical and pH Measurements*

Absorbance measurements were performed using a UV-2401PC/2501PC UV-visible spectrophotometer with a TCC-240A temperature-controlled holder (Shimadzu, Tokyo, Japan) at $(37.0 \pm 0.2)$ °C. The reaction was carried out in a 1 mL poly(methyl methacrylate), PMMA, cuvette for all the different input combinations. The increase of absorbance was monitored as a function of time, at λ = 420 nm, resulting from the formation of ferricyanide due to oxidation biocatalyzed by laccase. The gate time, 800 sec, was selected to obtain a well-defined **OR** function with filtering property. Absorbance was recalculated to the concentration of ferricyanide by using the Lambert-Beer law, where the molar extinction coefficient of ferricyanide[58] at λ = 420 nm is $\varepsilon_{420} = 1$ mM$^{-1}$·cm$^{-1}$ (thus resulting in numerically the same values for the absorbance and concentration of ferricyanide). The change of the pH in the system was simultaneously measured using a Mettler Toledo pH meter.

**REALIZATION OF THE OR GATE AND ITS KINETIC MODELING**

The present **OR** gate realization is described in Scheme 1. It involves two enzymatic processes and the buffering part. The first enzyme is esterase, of concentration to be denoted as $E(t)$, where $t$ is the time. Here esterase reacts with ethyl butyrate (concentration $A(t)$, Input 1) or methyl butyrate (concentration $B(t)$, Input 2), or both, biocatalyzing production of ethanol and methanol, respectively. Butyric acid (concentration $U(t)$) is a byproduct of the process, and its production lowers the pH of the system from its initial value of $pH(0) = 9.0$. The pH of the system varies from 9.0 to as low as 4.2, depending on the input concentrations and whether the buffer is present and its initial quantity.

Tris buffer (initially added as $T_0$ moles per unit volume) was introduced for "filtering" as explained shortly. The experiments were carried out with no buffering: $T_0 = 0$ mM, with buffering yielding high-quality **OR** gate: $T_0 = 4$ mM, and with excessive buffering which impairs the **OR** gate realization: $T_0 = 8$ mM. The added buffer instantaneously equilibrates to produce a protonated form $TH^+$. This hinders the reduction of the pH. However, this mild alkaline buffering effect persists only in the range of pH from 9 to slightly under 7.2, or until Tris is fully in the protonated form. Thus, changing the initial buffer supply, $T_0$, we can selectively delay the onset of the drop in pH. However, ultimately the pH will decrease due to the continuing production of butyric acid; $(pK_a)_{\text{butyric acid}} = 4.82$.[59]

As the pH in the system decreases, the biocatalytic action of laccase results in the conversion of $K_4Fe(CN)_6$ into $K_3Fe(CN)_6$, of concentrations to be denoted $F(t)$ and $P(t)$, respectively. The product of the process, $P(t_g)$, is measured at the gate time $t_g = 800$ sec by absorbance (denoted Abs) at $\lambda = 420$ nm. For the present system, Abs and $P$ are numerically the same (but have different units), as noted earlier.

For our model biochemical-logic gate system, we took the physical zeros, $A_0(0) = B_0(0) = 0$, of the signals as binary **0**s, and experimentally convenient input values $A_1(0) = 10$ mM and $B_1(0) = 10$ mM as logic **1**s. Logic **1** of the output, $P_1(t_g)$, is set by the gate function itself (see below). For the **OR** gate realization, we require the appropriate output values: **0**, **1**, **1**, **1**, at the four logic-input combinations, **00**, **01**, **10**, **11**, respectively. For study of the noise handling, however, behavior of the system response to the inputs should also be explored[4,5,60] near and between the logic input values. The latter study is carried out in terms of the scaled, binary-range (zero to one) variables,

$$x = A(0)/A_1(0), \quad y = B(0)/B_1(0), \quad z = P(t_g)/P_1(t_g), \qquad (1)$$

and the function $z(x,y)$. This is done in the next section. Here we focus on the (bio)chemical kinetics in terms of the original variables. Note that in practical applications the logic **1** values of the inputs will be determined[60] by the environment in which the gate is used, whereas the logic zero values of the inputs (and output) need not be at the physical zeros.[25,29,60]

Figure 1 (the figures are appended starting on page 18) shows the **OR** gate realization without filtering. As expected, the output values are at zero at the logic **00**, and are close to each other for logic **01**, **10**, **11** inputs, setting the binary **1** for the output. In the figure, the experimental data are shown as compared to a model calculation. The latter was based on kinetic process parameters taken from the literature, as detailed later, rather than of direct fitting of our data. Superficially, the system response shown in Figure 1 is suggestive of a realized **OR** logic. However, its noise-handling properties are not satisfactory as quantified in the next section. Figure 2 shows data and model results for a realization with the filtering processes added, for buffering at $T_0 = 4$ mM. Here not only is the visual **OR** logic realization convincing, but furthermore the response surface is sigmoid (flat in the vicinity all the logic points) offering a high-quality binary gate for noise handling. This is also quantified later. Finally, Figure 3 shows results for $T_0 = 8$ mM. In this case the "filtering" is excessive, which actually impairs the **OR** gate realization.

As described in earlier works,[25,36,37,40,60] evaluation of noise-handling properties and generally quality of the realized systems as "logic gates" for biochemical information processing requires a description of the system response in terms of the logic variable ranges, namely, as the function $z(x, y)$. Enzymatic kinetics, especially for coupled enzymatic cascades, is complicated and difficult to quantify experimentally and model theoretically. However, for the "logic function" analysis we only need an approximate, schematic description in terms of the key reaction steps to map out the shape of $z(x, y)$ near the four logic points, as well as to understand the control of the main features of this response-surface function by varying the "gate machinery" quantities. The latter refer to physical conditions of the experiments if these can be adjusted, and also to the initial concentrations of those chemicals which can be adjusted but do not constitute the inputs or output, for example, the two enzymes, or the filter process chemical(s), here, the buffer.

In the rest of this section we will summarize such a description for processes involved in the present system functioning. Interestingly, unlike all the other earlier studied enzymatic gate realizations, the present system parameters were all well-studied in the literature and therefore no actual data fitting was required. The response surfaces shown in Figures 1, 2, 3 are based on published reaction rates as detailed shortly. Note that Figures 1, 2, 3 show the actual concentrations (for the inputs) and measured optical signal (for the output) not scaled to logic-variable ranges in terms of $x, y, z$.

The biocatalytic function of esterase (concentration $[E] = E(t)$) can be described via the standard schemes

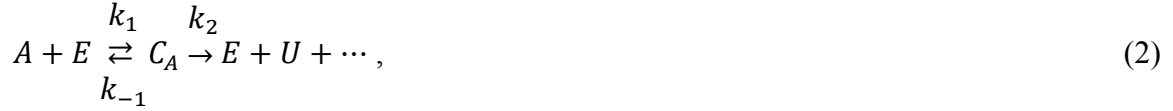

$$A + E \underset{k_{-1}}{\overset{k_1}{\rightleftarrows}} C_A \overset{k_2}{\rightarrow} E + U + \cdots, \qquad (2)$$

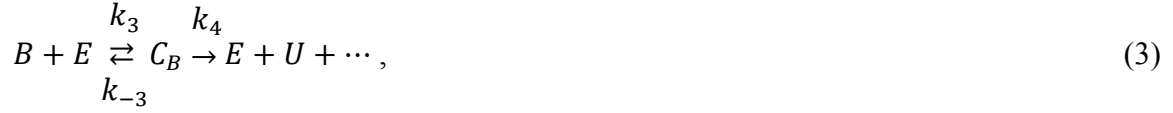

$$B + E \underset{k_{-3}}{\overset{k_3}{\rightleftarrows}} C_B \overset{k_4}{\rightarrow} E + U + \cdots, \qquad (3)$$

where $C_A$ and $C_B$ are (time-dependent) concentrations of complexes, and the notation for the other chemicals was defined earlier. Here we take $E(0) = 0.5$ U = $4.85 \times 10^{-5}$ mM, and of course $U(0) = 0$. Furthermore, the present system is in the regime of fast forward reaction in the first step, and therefore we can set $k_{-1}, k_{-3} = 0$. As the pH of the system varies over a wide range, the effective rate constants in Equations (2) and (3) will also vary, because of the acidification of the enzyme and complexes. As in the earlier work,[37] we described this pH dependence by assuming fast acidification equilibria, for example, $k_1 = \bar{k}_1 K_E / ([H^+](t) + K_E)$, with $\bar{k}_1 =$ 1.018 mM$^{-1}$sec$^{-1}$.[61] We also use[61] $\bar{k}_2 = 1.603$ sec$^{-1}$, $\bar{k}_3 = 0.639$ mM$^{-1}$sec$^{-1}$, $\bar{k}_4 = 3.990$ sec$^{-1}$. The acidification equilibrium constants,

$$K_E = [E][H^+]/[EH^+] = 0.0068 \text{ mM}, \qquad (4)$$

and similarly for $K_{C_A} = 0.026$ mM and $K_{C_B} = 0.039$ mM, were taken from the literature as well.[37,51,61] The effective rate constants thus become time-dependent via pH($t$). Next, we set up rate equations for the chemicals involved in the "esterase" part of the cascade. Here we only show one of these equations for illustration,

$$\frac{dC_A}{dt} = k_1(t)A(t)E(t) - k_2(t)C_A(t). \qquad (5)$$

Note that the kinetics of the "esterase" part is thus coupled to the other processes via the variation of pH with time. The buffer part of the kinetics is actually practically instantaneous. Standard charge balance equilibria equations were set up for the present system. We only outline pertinent points here. The conventional approximations for buffer functioning cannot be used because of the large range of pH variation, from basic pH(0) = 9.0, via neutral, to acidic down to about pH($t_g$) = 4.2, in the present experiments. Instead, the actual charge bookkeeping was done for all the ions in the system, and the charge-balance equation, as well as dissociation equilibrium equations for each of the relevant chemicals were included in the overall set of equations solved numerically. Specifically, for Tris buffer[62] we have $(pK_a)_{Tris} = 8.06$.

Consider now the processes biocatalyzed by laccase, the function of which can be schematically described as follows,[53]

$$4H^+ + L + 4F + O_2 \underset{k_{-5}}{\overset{k_5}{\rightleftarrows}} C_L \overset{k_6}{\rightarrow} L + 4P + 2H_2O. \qquad (6)$$

The notation for the concentration of various chemicals has been defined earlier, and the initial values are $L(0) = 0.5$ U $= 3.78 \times 10^{-4}$ mM, $F(0) = 1$ mM, $P(0) = 0$ mM. The actual mechanism of action of laccase is much more complicated than that schematically shown in Equation (6). It involves several steps and formation of more than one type of a complex (rather than our $C_L$). The full description of these processes (such as for instance consecutive capture/oxidation of ferrocyanide molecules) would require numerous rate constants. However, in the present regime it has been established in the literature[53] that an effective description is possible in terms of the lumped-step processes shown with their specific rates in Equation (6). In our regime of very low initial enzyme concentration, we can set $k_{-5} = 0$. For the other two effective rates we took the literature values, which are actually pH dependent and known phenomenologically from direct measurements. The pH dependence is bell-shaped and was phenomenologically fitted directly to the data. For reference, the maximal values are close to $\tilde{k}_5 = 1.6 \times 10^{-2}$ mM$^{-1}$sec$^{-1}$ for pH $\approx 8$, and close to $\tilde{k}_6 = 0.14$ sec$^{-1}$ for pH $\approx 7.5$. Here the rate constants are for the effective processes: $L + F \xrightarrow{\tilde{k}_5(pH)} C_L$, with the fact that four steps of capturing ferrocyanide are involved already absorbed in the effective rate constant, and similarly $C_L \xrightarrow{\tilde{k}_6(pH)} L + P$.

We can now write the remaining set of rate equations for the concentrations of chemicals involved, for example,

$$\frac{dL}{dt} = -\tilde{k}_5(\text{pH}(t))L(t)F(t) + \tilde{k}_6(\text{pH}(t))C_L(t). \tag{7}$$

With careful bookkeeping to account for instantaneous equilibration processes, e.g., Equation (4), we can then program the numerical calculation yielding the theoretical surfaces shown in Figures 1, 2, 3. Comparison of these results with the data, as well as the analysis of the quality of the realized **OR** gates are reported in the next section.

**QUALITY OF THE OR GATE REALIZATION AND DISCUSSION OF THE RESULTS**

To discuss the quality of the realized **OR** gate we consider signal processing by it in the context of noise handling. There could be several sources of inaccuracy in gate functioning affecting its usability as part of information and signal processing networks. The most obvious source of noise is the inaccuracy in the actual gate realization. This is apparent in Figure 3, where the realized response values for logic inputs that should yield output **1** are obviously not equal. The actual definition of the expected value or range of acceptable outputs corresponding to logic **1** may vary according to the application/networking of the gate. For definiteness, for our model system we choose the theoretically calculated value $P_1(t_g) = P(t = t_g; A_1(0), B_1(0))$, namely the output theoretically predicted within our specific model at inputs **11**. The average deviation of the actual model-predicted logic-**1** values at the three inputs **01**, **10**, **11**, from the selected reference **1** is given in Table 1 (see page 16) as percentage of the logic 0 to 1 value range.

Leaving aside the noise in the actual experimental data (which is another source of noise, discussed shortly), one can then ask why is filtering needed at all. According to the above criterion, the original no-filter gate has the least "inaccuracy" error. However, another source of noise has to be considered and is actually more important for network realizations,[28,60] for which analog and also (for larger networks) digital error correction has to be utilized, the latter via network design. The former, analog noise buildup can occur via amplification of the noise in the two input signals. Typically, biochemical signals are noisy, with few-percent noise levels at least. Input noise will be amplified or suppressed by the gate function when transmitted to the output. This transmission factor can be estimated for smooth gate-response surfaces (which is the case here) by calculating the absolute value of the gradient (the slope) in terms of the logic variables introduced earlier: $|\vec{\nabla} z(x,y)|$. Figure 4 gives a contour plot of the slope values for the non-filter gate. The calculated gradient near the logic **00** inputs is rather large: close to 690% amplification (see Table 1). Therefore, while visually it might look as a reasonable **OR** gate realization (Figure 1), the unfiltered system is not good for any networking in applications.

The properly filtered system (Figure 2) does not have this problem. Indeed, the added filtering process has made the noise transmission at all the four logic points into actual noise suppression, with the largest transmission factor, shown in Table 1, safely below 100%. It is approximately 14% for the 4 mM buffer concentration. Actually, the third system (Figure 3) has a mild noise amplification, about 113%. Indeed, the present mechanism for "filtering" only applies near zero output values. The oversupply of the buffer (8 mM) made the response too steep near the logic **11** inputs. Anyhow, we already noted that the 8 mM system is simply not an accurate **OR** gate function. The absolute values of the gradients (slopes) are mapped out in Figures 5 and 6. These values allow us to estimate the ranges of noise in the input signals off the fixed reference **0** and **1** values that the gate can "tolerate" because the deviations fall within the region of $|\vec{\nabla} z(x,y)| < 1$. These "noise tolerance" measures as percentages of the 0 to 1 logic-range span are also given in Table 1. The result is of course non-zero only for the 4 mM system.

Thus, filtering can improve the quality of the **OR** gate realization for networking applications. However, care must be exercised not to overdo it, in order to retain an accurate **OR** function. Furthermore, in most cases added filtering also causes signal intensity loss. This is clearly seen by comparing the maximal outputs (those near logic **11** inputs) in Figures 1, 2, 3. If we take the original (non-filtered) value as reference 100%, then with respect to it the percentage loss of intensity is as given in Table 1. One of the advantages of moderate filtering (4 mM) is that only 11% of the overall signal intensity is lost. Loss of signal intensity makes *all* sources of error and noise more significant on the relative scale of the signal strength.

Finally, the actual gate-function realization in each specific experimental setting can result not only in systematic deviations from the desired logic-point output values but also in random noise in the values of the output $P(t = t_g; A(0), B(0))$ or its logic-range equivalent $z(x,y)$. To estimate the noise level in the actual data measured as compared to the model predictions, we calculated the normalized mean-linear and root-mean-square (RMS) data-vs.-model deviations as averages over all the points taken for varying inputs, for each of the three gate realizations studied. These averages are given in Table 1 as percentages of the 0 to 1 logic-value output ranges. While the large values of the quadratic moments indicate that the present output data are rather noisy, the fact that the linear deviation moment is noticeably smaller than the quadratic moment confirms that the theoretical model works reasonably well.

In summary, we reported the first realization and study of a simple biocatalytic system with a double-sigmoid filter (sigmoid in both inputs) response offering the **OR** binary gate logic function. We also considered various gate-functioning quality criteria which can be used to select optimal realizations for future applications.

**Acknowledgements**

Research funding by the NSF, via awards CCF-1015983 and CBET-1066397, is gratefully acknowledged. Technical assistance of graduate students, M. A. Arugula, S. Chinnapareddy, S. Korkmaz in collecting the experimental data, and helpful discussions with Dr. Jan Halámek are acknowledged.

**Table 1: Quality Measures of Gate-Function Realization**

| $T_0$ (mM) | Spread of logic **1** outputs | Maximum noise transmission factor at the logic points | Input noise tolerance | Signal intensity loss | Average linear data-vs.-model deviation | RMS data-vs.-model deviation |
|---|---|---|---|---|---|---|
| 0 | 0.9% | 6.94 (at **00**) | 0% | 0% | 7.6% | 22% |
| 4 | 6.9% | 0.14 (at **10**) | 10.3% | 11% | 9.7% | 16% |
| 8 | 67.3% | 1.13 (at **11**) | 0% | 52% | 4.8% | 18% |

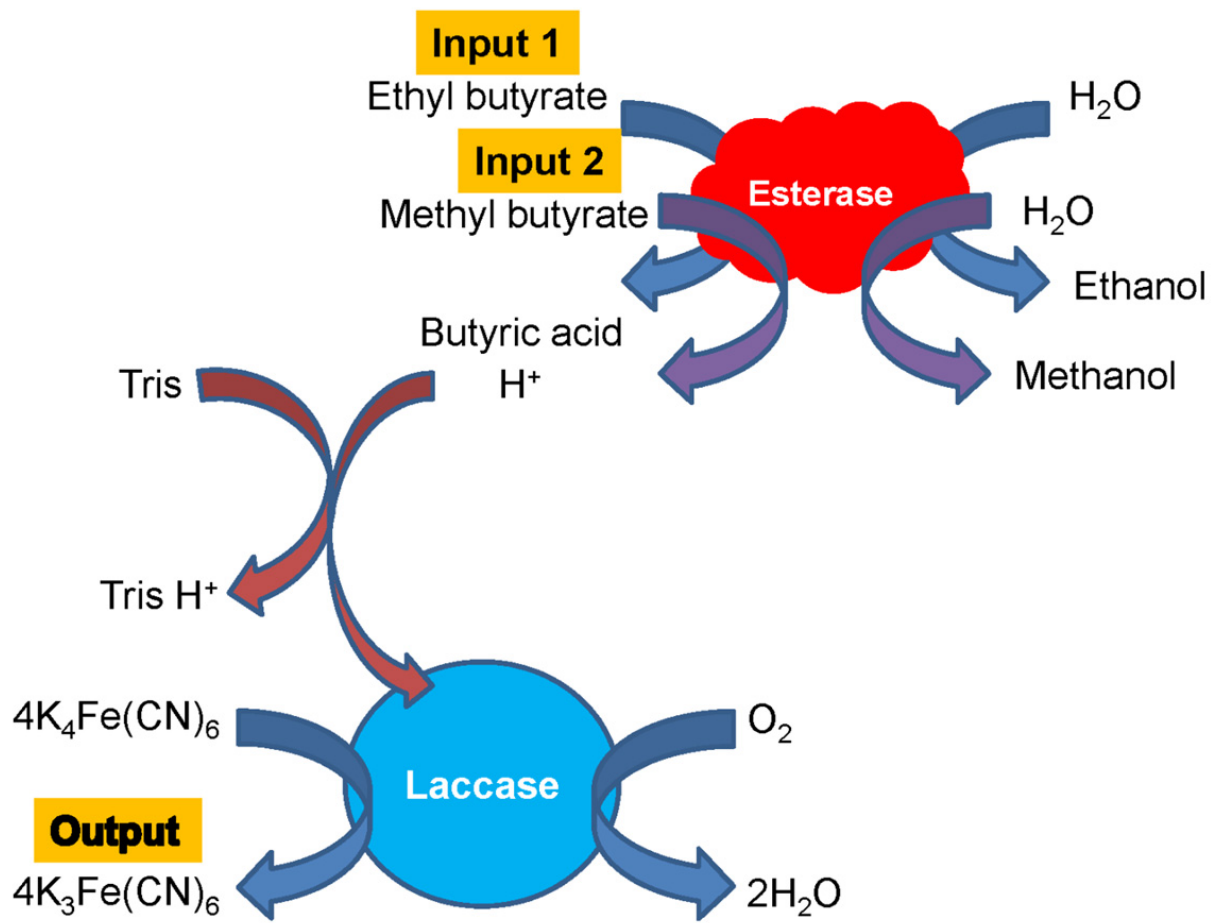

**Scheme 1.** Chemical and biochemical processes in the system.

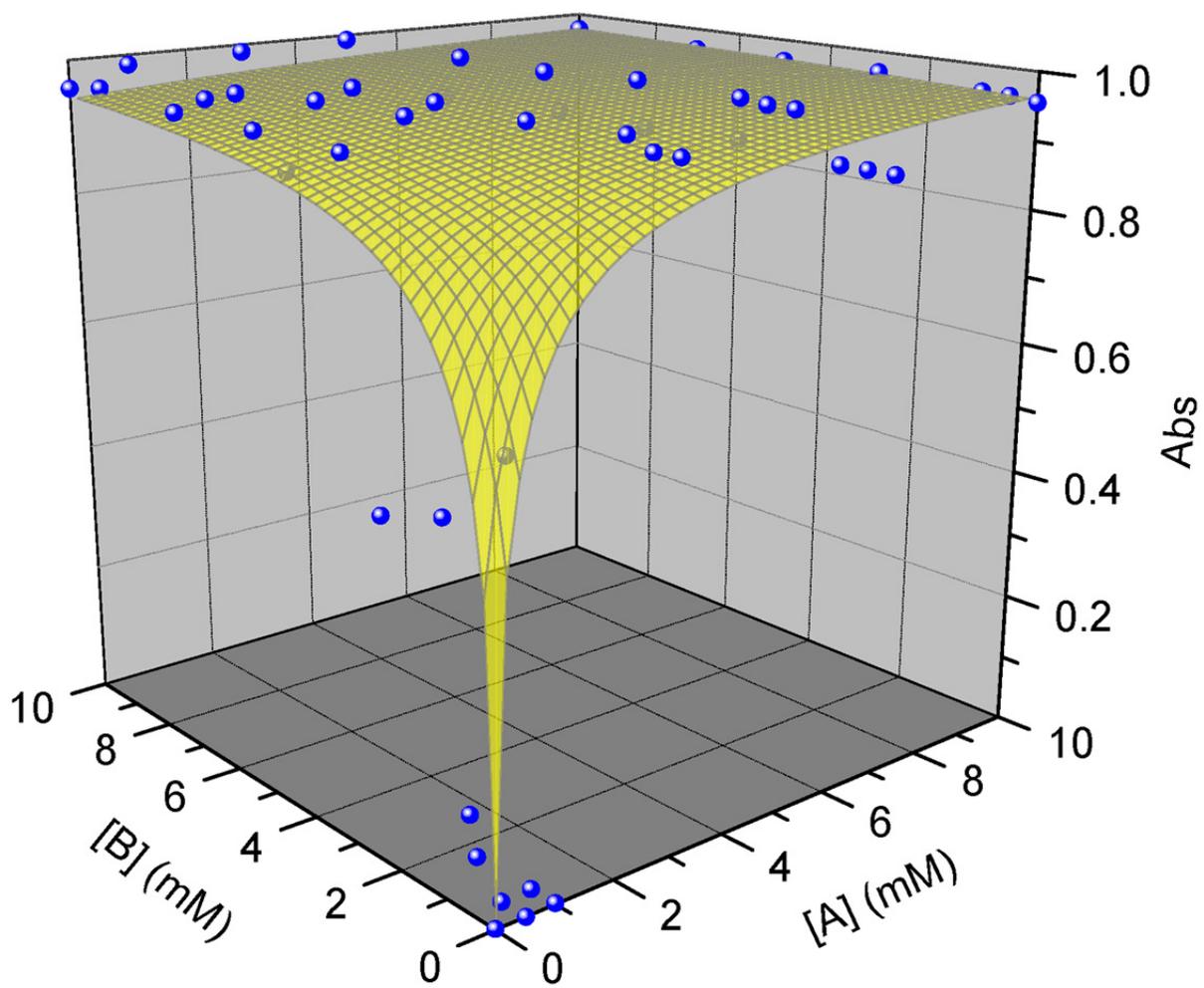

**Figure 1**. Spherical symbols show the 49 (7×7 grid of) experimental data points for the system without the buffer. These were measured at the gate time 800 sec, for various values of the two input concentrations (ethyl butyrate and methyl butyrate). The surface was calculated from the model described in the text.

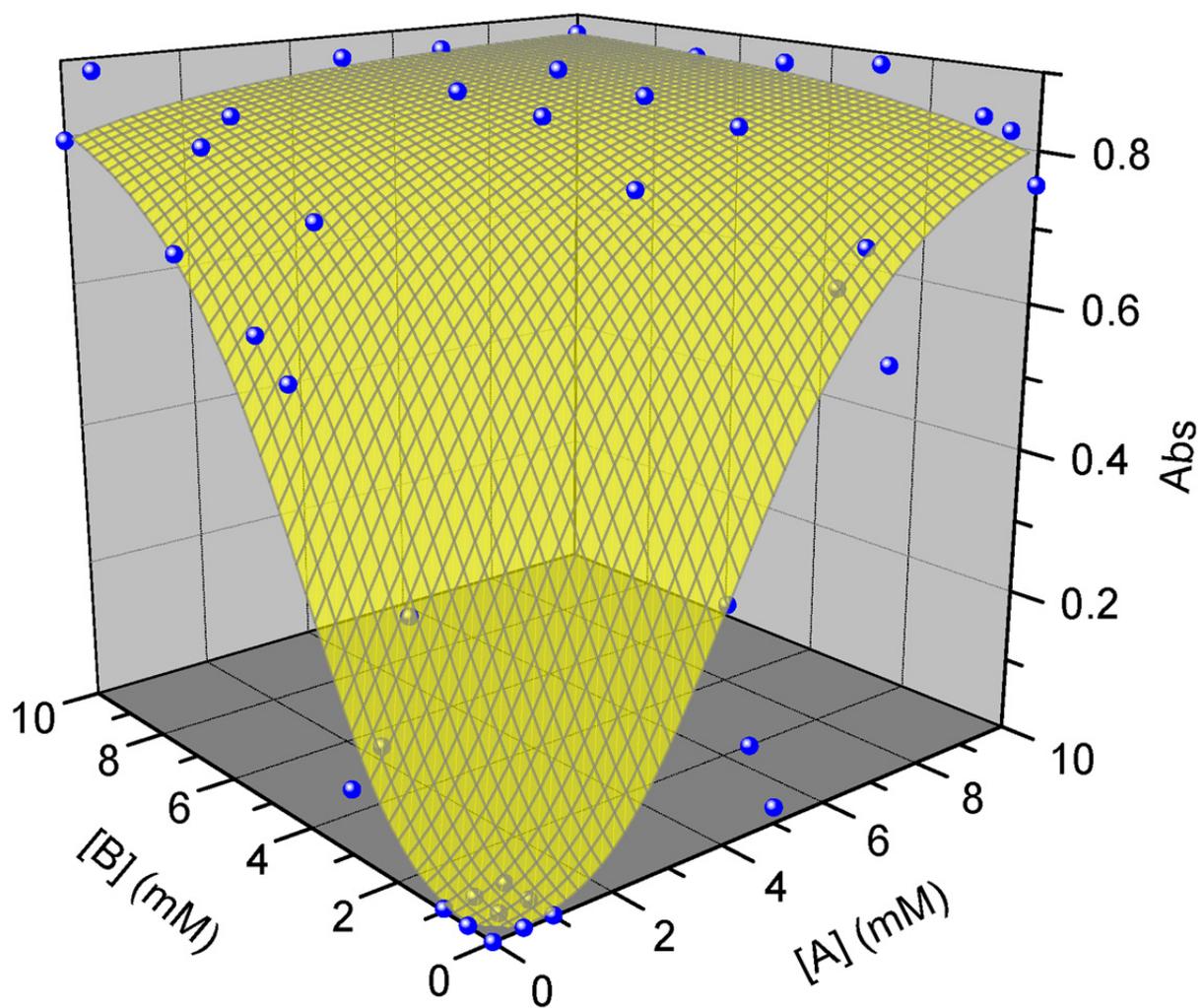

**Figure 2.** Experimental data and model-calculated surface (as in Figure 1) for the system with the 4 mM buffer added initially.

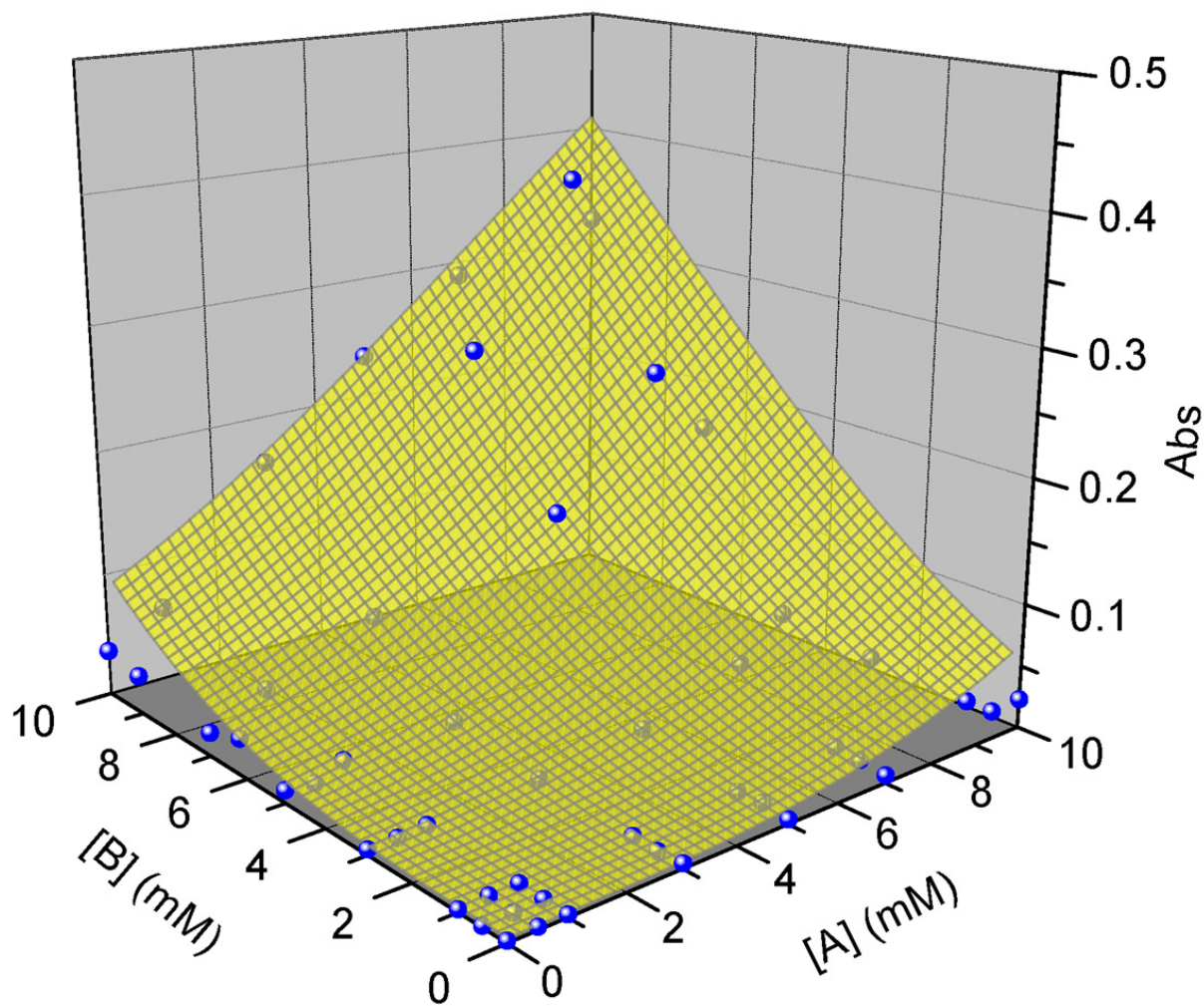

**Figure 3.** Experimental data and model-calculated surface (as in Figures 1 and 2) for the system with the 8 mM buffer added initially.

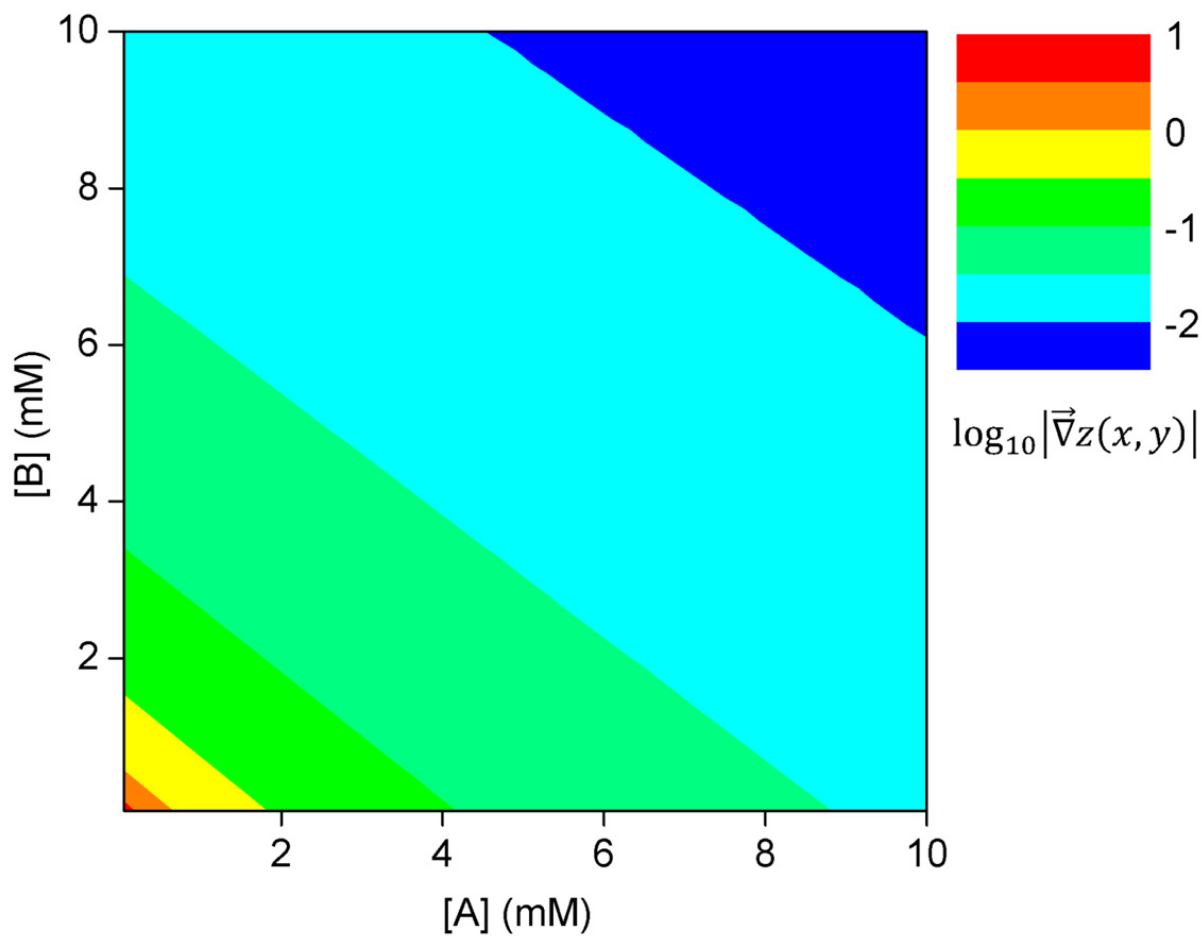

**Figure 4.** Color-coded contour plot of the absolute value of the gradient vector in terms of the rescaled variables, $z(x, y)$, calculated from the system without buffer, $T(0) = 0$ mM. Logarithmic scale is used for clarity.

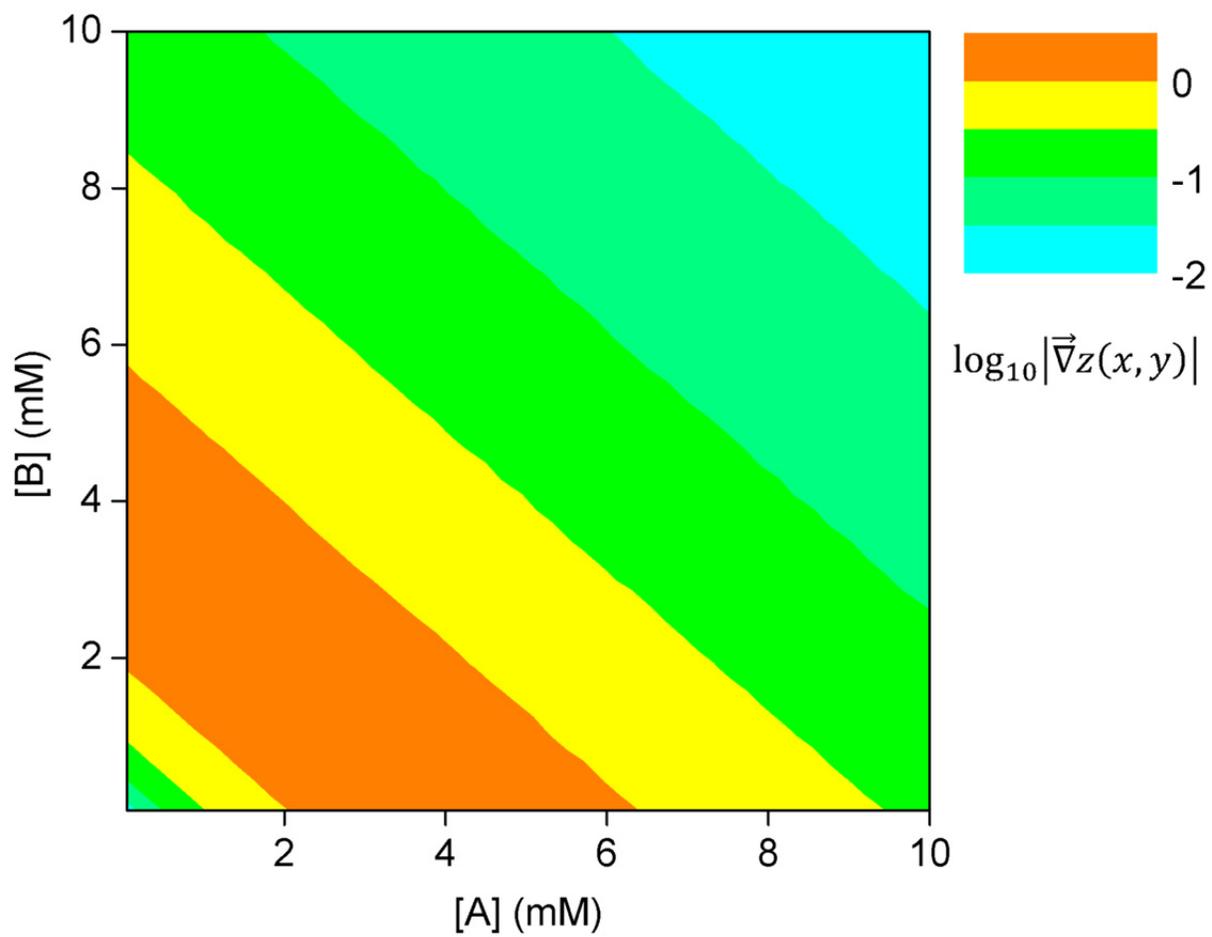

**Figure 5.** Color-coded contour plot of the absolute value of the gradient vector (as in Figure 4) for the system with the 4 mM buffer added initially.

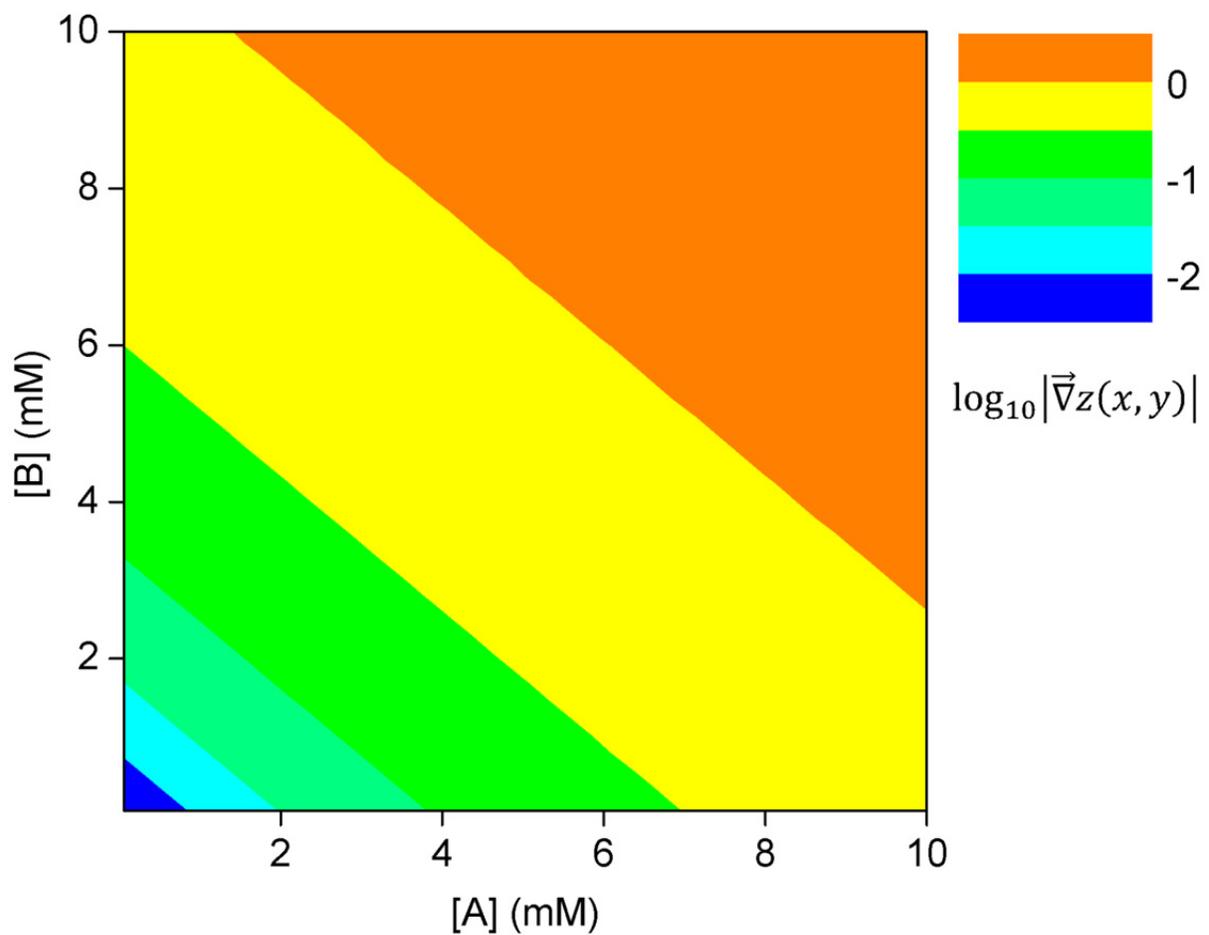

**Figure 6.** Color-coded contour plot of the absolute value of the gradient vector (as in Figures 4 and 5) for the system with the 8 mM buffer added initially.